\documentclass[%
 reprint,
 superscriptaddress,
 amsmath,amssymb,
 aps,
 prx,
 floatfix,
 longbibliography
]{revtex4-1}

\usepackage[%
	colorlinks=true,
	urlcolor=blue,
	linkcolor=blue,
	citecolor=blue,
	breaklinks=true
]{hyperref}
\usepackage[english]{babel} 
\usepackage{graphicx}
\usepackage{bm}
\usepackage{color}
\usepackage{siunitx} 
\usepackage[all]{hypcap} 

\begin{document}

\title{Resonant inelastic x-ray scattering of magnetic excitations under pressure}

\author{Matteo~Rossi}
\email{rossim@stanford.edu}
\altaffiliation[Present address: ]{Stanford Institute for Materials and Energy Sciences, SLAC National Accelerator Laboratory and Stanford University, 2575 Sand Hill Road, Menlo Park, California 94025, USA.}
\affiliation{ESRF -- The European Synchrotron, 71 Avenue des Martyrs, CS 40220, F-38043, Grenoble, France}

\author{Christian~Henriquet}
\affiliation{ESRF -- The European Synchrotron, 71 Avenue des Martyrs, CS 40220, F-38043, Grenoble, France}

\author{Jeroen~Jacobs}
\affiliation{ESRF -- The European Synchrotron, 71 Avenue des Martyrs, CS 40220, F-38043, Grenoble, France}

\author{Christian~Donnerer}
\affiliation{London Centre for Nanotechnology and Department of Physics and Astronomy, University College London, Gower Street, London WC1E6BT, United Kingdom}

\author{Stefano~Boseggia}
\affiliation{London Centre for Nanotechnology and Department of Physics and Astronomy, University College London, Gower Street, London WC1E6BT, United Kingdom}
\affiliation{Diamond Light Source, Harwell Science and Innovation Campus, Didcot, Oxfordshire OX11 0DE, United Kingdom}

\author{Ali~Al-Zein}
\altaffiliation[Present address: ]{Department of Physics, Faculty of Sciences, University of Balamand, P.O. Box 100, Tripoli, Lebanon.}
\affiliation{ESRF -- The European Synchrotron, 71 Avenue des Martyrs, CS 40220, F-38043, Grenoble, France}

\author{Roberto~Fumagalli}
\altaffiliation[Present address: ]{Dipartimento  di  Fisica,  Politecnico  di  Milano,  Piazza  Leonardo  da  Vinci  32, I-20133 Milano, Italy.}
\affiliation{ESRF -- The European Synchrotron, 71 Avenue des Martyrs, CS 40220, F-38043, Grenoble, France}

\author{Yi~Yao}
\altaffiliation[Present address: ]{Karlsruher Institut f\"ur Technologie, Institut f\"ur Festk\"orperphysik, Hermann-v.-Helmholtz-Platz 1, D-76344 Eggenstein-Leopoldshafen, Germany.}
\affiliation{ESRF -- The European Synchrotron, 71 Avenue des Martyrs, CS 40220, F-38043, Grenoble, France}

\author{James~G.~Vale}
\affiliation{London Centre for Nanotechnology and Department of Physics and Astronomy, University College London, Gower Street, London WC1E6BT, United Kingdom}
\affiliation{Laboratory for Quantum Magnetism, Ecole Polytechnique F\'ed\'erale de Lausanne (EPFL), CH-1015, Switzerland}

\author{Emily~C.~Hunter}
\affiliation{Centre for Science at Extreme Conditions, The University of Edinburgh, Mayfield Road, Edinburgh EH9 3JZ, United Kingdom}

\author{Robin~S.~Perry}
\affiliation{London Centre for Nanotechnology and Department of Physics and Astronomy, University College London, Gower Street, London WC1E6BT, United Kingdom}
\affiliation{Centre for Science at Extreme Conditions, The University of Edinburgh, Mayfield Road, Edinburgh EH9 3JZ, United Kingdom}

\author{Innokenty~Kantor}
\altaffiliation[Present address: ]{MAX IV Laboratory, Lund University, SE-221 00 Lund, Sweden.}
\affiliation{ESRF -- The European Synchrotron, 71 Avenue des Martyrs, CS 40220, F-38043, Grenoble, France}

\author{Gaston~Garbarino}
\affiliation{ESRF -- The European Synchrotron, 71 Avenue des Martyrs, CS 40220, F-38043, Grenoble, France}

\author{Wilson~Crichton}
\affiliation{ESRF -- The European Synchrotron, 71 Avenue des Martyrs, CS 40220, F-38043, Grenoble, France}

\author{Giulio~Monaco}
\altaffiliation[Present address: ]{Dipartimento di Fisica, Universit\`a di Trento, via Sommarive 14, I-38123 Povo (TN), Italy.}
\affiliation{ESRF -- The European Synchrotron, 71 Avenue des Martyrs, CS 40220, F-38043, Grenoble, France}

\author{Desmond~F.~McMorrow}
\affiliation{London Centre for Nanotechnology and Department of Physics and Astronomy, University College London, Gower Street, London WC1E6BT, United Kingdom}

\author{Michael~Krisch}
\affiliation{ESRF -- The European Synchrotron, 71 Avenue des Martyrs, CS 40220, F-38043, Grenoble, France}

\author{Marco~Moretti~Sala}
\email{marco.moretti@polimi.it}
\altaffiliation[Present address: ]{Dipartimento  di  Fisica,  Politecnico  di  Milano,  Piazza  Leonardo  da  Vinci  32, I-20133 Milano, Italy.}
\affiliation{ESRF -- The European Synchrotron, 71 Avenue des Martyrs, CS 40220, F-38043, Grenoble, France}

\begin{abstract}
Resonant inelastic x-ray scattering (RIXS) is an extremely valuable tool for the study of elementary, including magnetic, excitations in matter. Latest developments of this technique mostly aimed at improving the energy resolution and performing polarization analysis of the scattered radiation, with a great impact on the interpretation and applicability of RIXS. Instead, this article focuses on the sample environment and presents a setup for high-pressure low-temperature RIXS measurements of low-energy excitations. The feasibility of these experiments is proved by probing the magnetic excitations of the bilayer iridate Sr$_3$Ir$_2$O$_7$ at pressures up to 12 GPa. 
\end{abstract}

\maketitle


\section{Introduction}
\label{sec:intro}

Resonant inelastic x-ray scattering (RIXS) is a photon-in photon-out spectroscopic technique that allows to study the low-energy physics of materials by probing their elementary excitations \cite{Schuelke2007,Ament2011}. The RIXS process consists in the resonant photoexcitation of a core electron into an empty state above the Fermi level and the subsequent radiative de-excitation of the system. The final state of the system -- either the ground state itself or an excited state -- is characterized by the energy, momentum and polarization difference between the incident and emitted photons. Due to the resonant nature of the scattering process and to the use of x-rays, RIXS is an element- and orbital- selective, momentum resolved, bulk sensitive technique \cite{Ament2011}. Different types of excitations can be accessed over an extended energy range, from several hundreds of eV down to a few meV, and this limit is continuously being pushed. Indeed, in the last decade, enormous progresses have been made in terms of energy resolution by building RIXS spectrometers on dedicated beamlines \cite{MorettiSala2013,Lai2014,Dvorak2016,MorettiSala2018,Kim2018,Brookes2018}. At the same time, initial difficulties in the interpretation of RIXS experiments have been overcome by exploiting its complementarity to other experimental techniques and by developing suitable theories \cite{Ament2009,Haverkort2010,Ament2011,Ament2011b,MorettiSala2014}. A prominent example of RIXS application originates from the theoretical demonstration \cite{Ament2009,Haverkort2010,MorettiSala2011} and experimental observation \cite{Braicovich2010} that RIXS can detect single spin-flip excitations in copper-based superconductors (cuprates), leading to an extensive use of RIXS for the study of their magnetic dynamics \cite{LeTacon2011,Dean2013,Dean2015}, in a way complementary to inelastic neutron scattering (INS) \cite{Tranquada1989,Coldea2001,Headings2010}.

In the hard-x-ray energy range, RIXS benefits from an additional advantage: it can be coupled to complex sample environments to study matter under extreme conditions \cite{Rueff2010,Kim2016a}. In particular, the application of high pressure is a valuable tool to investigate the properties of matter \cite{Mao2018}, since it can effectively alter the electron density and, ultimately, induce structural, electronic and magnetic phase transitions \cite{Klotz2012,Gorkov2018,Mao2018}. However, high-energy-resolution RIXS measurements at high pressure are difficult \cite{Rueff2010,Kim2016a} and even more so when dealing with low-energy excitations, such as magnetic ones. As a matter of fact, to the best of our knowledge, no such measurements have ever been reported. Note that this field is rather unexplored as a whole, as the main probe of magnetic excitations, INS, is limited to relatively low pressures ($\sim 1$ GPa) \cite{Klotz2012}  by the competing requirements of the large sample volume required by neutron techniques ($\sim$ 1 cm$^3$) and high-pressure experiments.

We here discuss recent instrumental developments made at beamline ID20 of the ESRF -- The European Synchrotron (Grenoble, France) that allowed us to perform RIXS measurements of low-energy magnetic excitations under pressure. As a test case, we consider iridium oxides as they attracted the attention of the scientific community in the last decade \cite{WitczakKrempa2014,Rau2016}. Indeed, they feature electron-electron and spin-orbit interactions of comparable strength, leading to unexpected properties, including the so-called spin-orbit-induced Mott insulating state \cite{Kim2008,Kim2009}. The single-layer Sr$_2$IrO$_4$ ($n=1$) is the prototypical spin-orbit-induced Mott insulator \cite{Kim2008,Kim2009} and develops long-range (canted) antiferromagnetic (AFM) order below a N\'{e}el temperature of $\approx 240$ K \cite{Cao1998}. It has been in the focus of numerous theoretical and experimental investigations regarding its similarities with cuprates \cite{Kim2012,Kim2014a,DeLaTorre2015,Kim2015,Yan2015,Gretarsson2016a,Liu2016,Terashima2017,Pincini2017,Calder2018} and the possibility to host unconventional superconductivity \cite{Wang2011,Watanabe2013,Yang2014,Meng2014}. The bilayer Sr$_3$Ir$_2$O$_7$ is the $n=2$ member of the Ruddlesden-Popper series Sr$_{n+1}$Ir$_n$O$_{3n+1}$ and shares with Sr$_2$IrO$_4$ many physical properties, including a Mott-insulating ground state of relativistic nature \cite{Moon2008} and the tendency to develop long-range AFM order below $\approx 280$ K \cite{Cao2002,Boseggia2012a}. However, the microscopic mechanisms that govern magnetism in Sr$_3$Ir$_2$O$_7$ are debated since two different theoretical models have been proposed to interpret its magnetic structure and dynamics \cite{Kim2012a,MorettiSala2015}. The two models are mutually exclusive, as they cast the main exchange couplings into two opposite scenarios: Kim \emph{et al.} proposed a linear spin-wave approach with dominant intralayer interaction \cite{Kim2012a}, whereas Moretti Sala \emph{et al.} pursued a bond-operator mean-field approach with predominant interlayer, dimer-like interactions \cite{MorettiSala2015}. Despite fundamental differences, the two theoretical approaches fit equally well the magnetic dynamics of Sr$_3$Ir$_2$O$_7$ at ambient pressure, with a bandwidth of $\approx 70$ meV and an anomalously-large magnetic gap of $\approx 90$ meV. In order to discriminate between the two models, the evolution of magnetic excitations should be tested against well-defined and controlled perturbations of the system. In this respect, the magnetic dynamics of (Sr$_{1-x}$La$_x$)$_3$Ir$_2$O$_7$ was studied as a function of doping (up to $x=0.1$), but without conclusive evidence in favour of either model \cite{Hogan2016,Lu2017}. In addition, it is often argued that doping is not a ``clean'' perturbation, as it introduces disorder in the system, thus making the interpretation of experimental results less reliable. We therefore propose to use physical pressure as an alternative and clean manner to perturb the system and prove the feasibility of this approach. Indeed, the application of physical pressure does not modify the chemical composition of the system and can be gradually tuned from arbitrarily small to very large values, up to hundreds of GPa's. 

Thus, once established as a viable technique, it is not difficult to imagine that high-pressure RIXS to probe the magnetic quasiparticle spectrum will find general use and produce results of widespread interest.

\section{Background considerations}

High-pressure experiments require the use of so-called diamond anvil cells (DACs), in which two diamonds are pushed one against the other to apply pressure to a sample in between them. The sample chamber is obtained by perforating a small hole in a gasket that prevents the two diamonds from touching. The high-pressure environment therefore completely embeds the sample and poses several constraints on the experiment, including the limitation of x-ray experiments under high pressure to the hard x-ray range. RIXS excitations of single crystalline materials are usually probed at specific transferred momenta, hence the requirement of precise scattering geometries. However, the severe geometrical constraints imposed by the DAC and the gasket often limit the accessible scattering geometries. Another important aspect to take into account is that the sample environment absorbs part of the radiation. Since RIXS is a photon hungry technique, it is therefore imperative to minimize the x-ray absorption through the sample environment, as well as maximize the collected solid angle and optimize the sample volume. This last point is very delicate: on the one hand, the maximum reachable pressure  depends on the diamond culets, which ultimately limit the sample dimension; on the other hand, the sample size should exceed the x-ray penetration depth in order to maximize the RIXS scattering cross-section. The best compromise is that the sample has roughly the dimensions of the x-ray focal spot size ($\sim 20\times 10$ \SI{}{\micro\meter}$^2$) times the x-ray penetration depth ($\sim 8$ \SI{}{\micro\meter} in Sr$_3$Ir$_2$O$_7$ at the Ir $L_3$ edge). Finally, x-ray scattering from the sample environment -- typically non-resonant elastic and inelastic scattering from the gasket and/or the diamonds of the DAC -- contributes to the background noise, which may even be stronger than the actual RIXS signal. Note that elastically-scattered x-rays from a source slightly off the sample position will impinge on the crystal analyzer at an angle different than the nominal Bragg angle and will be interpreted as inelastic signal, thus potentially giving rise to artifacts in the RIXS spectrum. In addition, x-ray scattering from crystalline diamonds will feature sharp phonon peaks at an energy loss up to $\approx 170$ meV \cite{Warren1967}. Similarly, the gasket is also a source of inelastic scattering. In the case that amorphous beryllium is used, however, the phonon signal is a broad weak distribution, resembling the phonon density of states and extending up to $\approx 100$ meV \cite{Schmunk1962,Stedman1976}.

In order to cope with most of the difficulties described above, we adopted a scattering geometry as close as possible to \SI{90}{\degree} where the x-rays travel through the gasket of the DAC. For the horizontally polarized light used in our experiments, this geometry leads to a large suppression of all spurious processes arising from Thomson scattering. The choice of the gasket material is then restricted to beryllium, because of its low absorption of hard x-rays. The main drawbacks are the low shear strength and brittleness, which limit the maximum reachable pressure.

\section{Sample environment}

\subsection{Diamond anvil cell and gasket}
\begin{figure}
	\centering
	\includegraphics[width=\columnwidth]{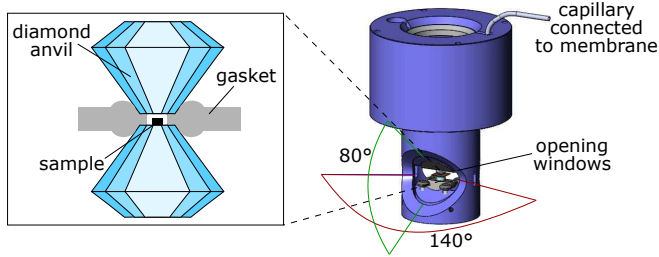}
	\caption{\label{fig:pano_DAC}Panoramic diamond anvil cell used for high-pressure RIXS experiments. The two diamonds are in the middle of the opening window. The compression axis is vertical and the scattering plane is horizontal in the Figure reference frame.}
\end{figure}

Diamond anvil cells are standard tools to perform experiments at pressures up to several hundreds of GPa \cite{Jayaraman1983,Bassett2009}. For our experiment we designed and built a customized DAC, shown in Fig.~\ref{fig:pano_DAC}: It features two large opening windows of \SI{140}{\degree} in the scattering plane and \SI{80}{\degree} in the plane orthogonal to it (see also Ref.~\onlinecite{Sahle2017}).

The extra-high 16-sided (100)-oriented diamond anvils are in the middle of the opening window. They are based on the type Ia Boehler-Almax design \cite{Boehler2004} (diameter of 3.1 mm, height of 2.72 mm, angle of the diamond faces of \SI{25}{\degree}). Diamonds of culet size ranging from 0.25 mm (maximum pressure $\sim 80$ GPa) to 0.5 mm (maximum pressure $\sim 30$ GPa) are used to cover different pressure ranges. A cylinder-piston system (modified LeToullec design \cite{Letoullec1988}) allows the movement of the diamond anvils along the compression axis. The anvils are pushed together by the application of force on one diamond through a gas-driven membrane that is connected to a pressure controller. The beryllium gasket with an outer diameter of 5 mm and a thickness of 0.2 mm was pre-indented to approximately 1/10 of the culet size (i.e. $\sim 30$-\SI{50}{\micro\meter}) before the experiment and a sample chamber with diameter approximately equal to 50-60\% of the culet size is laser-drilled in the center to accommodate the sample.

\begin{figure}
	\centering
	\includegraphics[width=0.8\columnwidth]{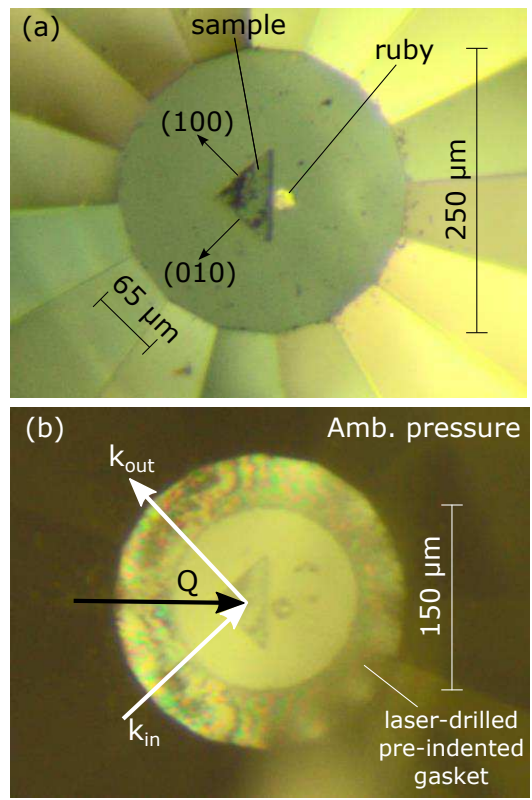}
	\caption{\label{fig:sample_loading_Sr327}(a) Sr$_3$Ir$_2$O$_7$ single crystal and ruby sphere loaded at the center of the culet with diameter of \SI{250}{\micro\meter}. The sample has a triangular shape with edges of $\approx \SI{65}{\micro\meter}$ parallel to high-symmetry directions of the lattice. (b) Sr$_3$Ir$_2$O$_7$ and ruby sphere after the pre-indented laser-drilled gasket has been placed between the diamond anvils. A sketch of the experimental geometry is also shown.}
\end{figure}

An image of a typical sample loading is shown in Fig.~\ref{fig:sample_loading_Sr327}. Panel (a) shows the culet of one anvil, in the middle of which a single crystal of Sr$_3$Ir$_2$O$_7$ is loaded. The sample has been polished roughly to the desired size. Sr$_3$Ir$_2$O$_7$ naturally cleaves into layers orthogonal to the $c$ axis, with sharp edges parallel either to the (100) or to the (110) axes, as illustrated in the Figure. We chose inert neon as pressure-transmitting medium since it is almost transparent to x-rays and it ensures a quasi-hydrostatic compression of the sample. Pressure in the DAC was determined by tracking the energy position of the $R_1$ fluorescence line of a ruby (Al$_2$O$_3$:Cr$^{3+}$) sphere placed next to the sample \cite{Mao1976}. Panel (b) shows the sample and the ruby sphere after the gasket has been put between the diamond anvils. The laser-drilled sample chamber with diameter of \SI{150}{\micro\meter} is visible.

\subsection{Cryostat}

\begin{figure*}
	\centering
	\includegraphics[width=0.6\textwidth]{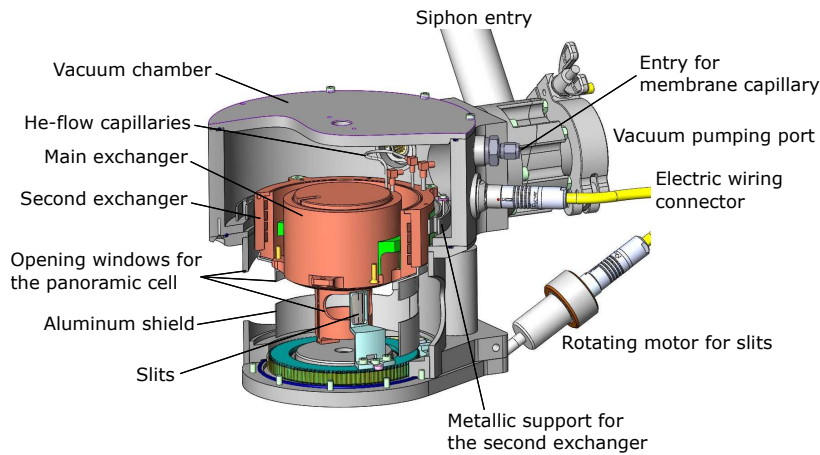}
	\caption{\label{fig:cryo_HP}Helium-flow cryostat for high-pressure low-temperature RIXS experiments. The panoramic DAC is lodged inside the main heat exchanger.}
\end{figure*}

The Panoramic DAC is placed inside a cryostat, in such a way that the compression axis is vertical and the wide opening windows are horizontal in the laboratory reference frame (Fig.~\ref{fig:cryo_HP}). The outer part of the cryostat is the vacuum chamber. It is made of stainless steel to withstand ambient pressure and is fixed on the RIXS sample goniometer. Two large windows have been opened in the walls of the chamber and covered by an x-ray-transparent 0.1-mm-thick Kapton foil. The vacuum is as good as $\le 10^{-7}$ mbar during the measurements. The vacuum chamber hosts the entries for the siphon, the vacuum pumping port, the electric wiring connectors, and a feed-through for the capillary that connects the pressure controller to the cell membrane. The Panoramic DAC is placed inside a copper main heat exchanger, which is cooled down by a continuous flow of helium. Its temperature is measured by a Cernox\textregistered~sensor and regulated by resistive Kapton foil heaters. Helium then flows through stainless steel capillaries into a copper secondary heat exchanger before it is evacuated through the input siphon. The temperature of the secondary exchanger is measured and regulated in a similar fashion as the main heat exchanger. It is designed such that the contact surface between the secondary exchanger and the vacuum chamber is minimized and the thermal paths are maximized for thermal decoupling. Furthermore, thermally-insulating washers are utilized to fix the two elements. Because of the cylindrically-symmetric design of the cryostat, the position of the DAC moves by less than $\approx$ \SI{40}{\micro\meter} in the horizontal plane and by $\approx$ \SI{160}{\micro\meter} along the vertical axis when cooled from room temperature to 100 K. 

\begin{figure}
	\centering
	\includegraphics[width=\columnwidth]{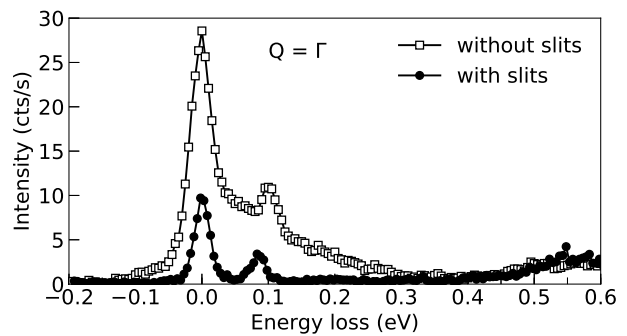}
	\caption{\label{fig:slits}RIXS spectra of Sr$_3$Ir$_2$O$_7$ collected at equivalent $\Gamma$ points with (filled circles) and without (empty squares) slits. Spectra are plotted in counts per second.}
\end{figure}

A system of motorized slits is placed inside the cryostat chamber. The slits rotate around the DAC compression axis at a distance of 16.5 mm from the sample. The slit aperture defines the region around the sample that is seen by the spectrometer, therefore their use allows to remove most of the background coming from the sample environment \cite{Yoshida2014}. The effect of the slit is clear upon inspection of Fig.~\ref{fig:slits}, which shows RIXS spectra of Sr$_3$Ir$_2$O$_7$ measured with (filled circles) and without (empty squares) slits. The background due to the sample environment extends up to energy losses of $\sim 0.3$ eV and almost covers the magnetic signal at $\sim 0.1$ eV. The use of slits suppresses most of the background and allows to safely recover the magnetic signal from the sample. Three slit apertures are available: 0.08 mm, 0.16 mm and 0.32 mm. The field of view of the spectrometer is 0.2 mm, 0.28 mm and 0.45 mm in the three cases, respectively, considering a mask of 15 mm placed in front of the crystal analyzer. The field of view is related to the diamond culet size and sample dimension and eventually to the pressure range of interest. Hence, the smallest slit aperture is used when pressures in the $\sim 10$ GPa range are targeted, while larger apertures are used for pressures in the MPa to $\sim 1$ GPa regime. In this work, all spectra are collected with the smallest slit aperture.

\section{RIXS measurements}

Iridium $L_3$ edge ($\approx 11.22$ keV) RIXS measurements were performed at beamline ID20 of the ESRF. The beamline was designed and constructed  within the framework  of the recent ESRF Upgrade -- Phase I. It is a flagship beamline for resonant and non-resonant inelastic hard-x-ray scattering spectroscopy \cite{MorettiSala2018,Huotari2017}. It has been conceived to maximize the photon flux at the sample position, cover a wide energy range (4-20 keV), optimize the energy resolution and guarantee a small stable beam over a long interval of time. These are essential requirements for RIXS experiments, in particular under extreme conditions. During the measurements, the incoming x-rays were monochromated to 11.2165 keV by the joint use of a Si(111) double-crystal monochromator and a backscattering Si(844) post-monochromator. A Kirkpatrick-Baez (KB) mirror system focused the incoming beam to a spot size smaller than $10\times\SI{20}{\micro\meter}^2$ (vertical $\times$ horizontal). The x-rays scattered by the sample were energy-analyzed by a Rowland-circle-based spectrometer, equipped with a spherical Si(844) diced crystal analyzer and a position-sensitive detector. The overall energy resolution is as good as 25 meV and the incident photon flux is $\sim10^{12}$ photons/s at the sample position \cite{MorettiSala2013,MorettiSala2018}. Further information about the RIXS spectrometer of beamline ID20 can be found in Ref.~\onlinecite{MorettiSala2018}.

A sketch of the scattering geometry employed during the measurements is shown in Fig.~\ref{fig:sample_loading_Sr327}(b). The incoming and outgoing photon wave vectors are $\mathbf{k}_\mathrm{in}$ and $\mathbf{k}_\mathrm{out}$, respectively, while $\mathbf{Q} = \mathbf{k}_\mathrm{in} - \mathbf{k}_\mathrm{out}$ is the momentum transfer. Note that this geometry allows to reach the boundary $(\pi, \pi)$ of the two-dimensional Brillouin zone.

\section{Results and discussion}

\begin{figure}
	\centering
	\includegraphics[width=\columnwidth]{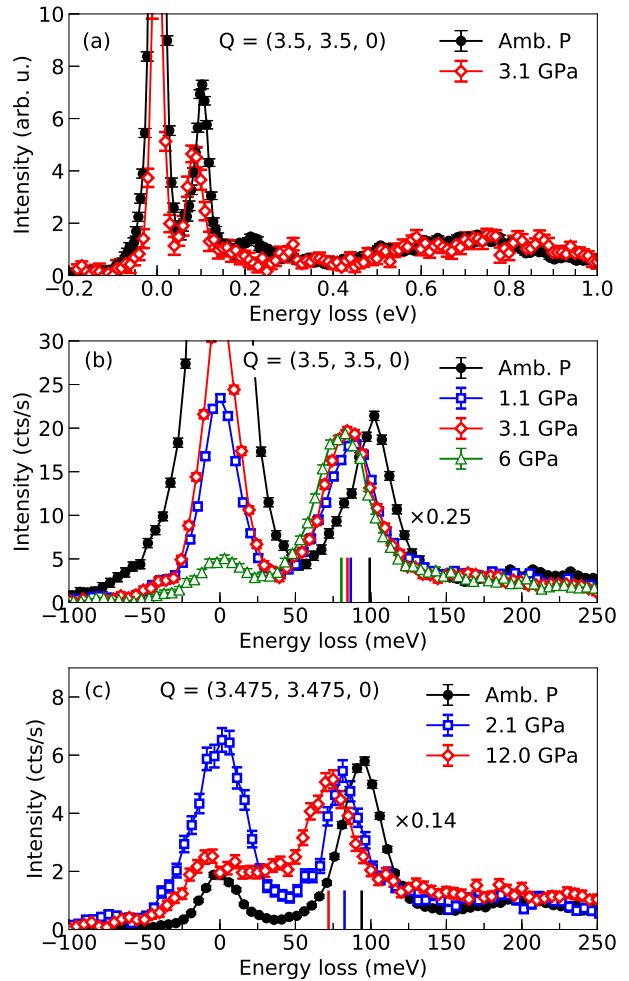}
	\caption{\label{fig:pressure_dependence}(a) Iridium $L_3$-edge RIXS spectra of Sr$_3$Ir$_2$O$_7$ measured at ambient pressure (black circles) and at $P = 3.1$ GPa (red diamonds). The spectra were collected at $T = 150$ K and $\mathbf{Q} = (3.5, 3.5, 0)$ r.l.u. and are normalized to the orbital excitations above 0.4 eV. (b) Magnetic region of the RIXS spectra measured at $T = 150$ K and $\mathbf{Q} = (3.5, 3.5, 0)$ r.l.u. showing the softening of the magnetic mode with pressure. (b) Magnetic region of the RIXS spectra measured at $T = 100$ K and $\mathbf{Q} = (3.475, 3.475, 0)$ r.l.u. The spectra acquired at ambient pressure in panels (b) and (c) are multiplied by 0.25 and 0.14, respectively, in order to compare them to the high-pressure RIXS data.}
\end{figure}

The setup described above has been employed to determine the pressure dependence of the magnetic structure and dynamics of Sr$_3$Ir$_2$O$_7$. Figure~\ref{fig:pressure_dependence}(a) displays overview raw RIXS spectra of Sr$_3$Ir$_2$O$_7$ measured at momentum transfer $\mathbf{Q} = (3.5, 3.5, 0)$ r.l.u. corresponding to the two-dimensional Brillouin zone boundary ($\pi$, $\pi$). The spectra are normalized to the features above 0.4 eV. As can be seen, both spectra measured at ambient pressure (black circles) and at $P = 3.1$ GPa (red diamonds) are qualitatively similar: besides the elastic line, both spectra feature a sharp (FWHM $\approx 0.04$ eV) peak at $\sim 0.1$ eV and a broad (FWHM $\approx 0.45$ eV) excitation centered at $\sim 0.75$ eV. The first feature is ascribed to the single magnon excitation \cite{Kim2012a,MorettiSala2015}, while the second feature is ascribed to the transition of a hole from the $j_\mathrm{eff}=1/2$ ground state to the $j_\mathrm{eff}=3/2$ band, in agreement with previous works on the sister compound Sr$_2$IrO$_4$ \cite{Kim2012,Kim2014}. The pressure dependence at room temperature of the latter excitation has been already investigated by Ding \emph{et al.}, who have found that the peak position changes by at most $\sim 15$\% from ambient pressure to approximately 65 GPa, while the width does not undergo a significant change \cite{Ding2016}. We note that magnetic excitations could not be resolved in the RIXS spectra measured by Ding \emph{et al.} \cite{Ding2016} not only because the measurements were carried out at room temperature, but also because of the spurious signal from the sample environment extending up to several hundreds of meV. Instead, magnetic excitations are clearly distinguished in our spectra, as highlighted in Fig.~\ref{fig:pressure_dependence}(b) and Fig.~\ref{fig:pressure_dependence}(c), and show an unambiguous softening when pressure is increased up to 12 GPa. In Fig.~\ref{fig:pressure_dependence}(b,c), the spectra are plotted in counts per second (cts/s) and the spectrum collected at ambient pressure is multiplied by a factor for better comparison. We observe that the intensity drops by a factor $\sim 4$-7 when the high-pressure sample environment is employed.

\begin{figure}
	\centering
	\includegraphics[width=\columnwidth]{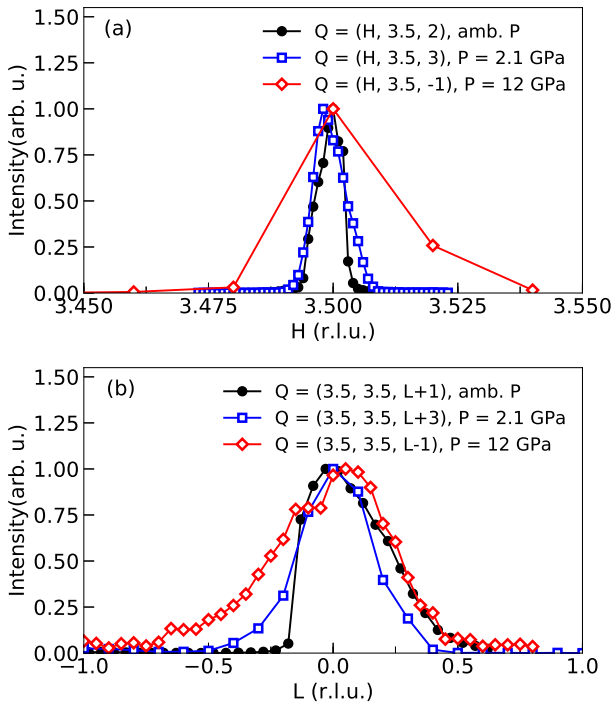}
	\caption{\label{fig:HKL_scan}Pressure dependence of the magnetic diffraction peaks of Sr$_3$Ir$_2$O$_7$ measured by collecting the elastic signal while scanning along high-symmetry directions of the reciprocal lattice parallel to the (100) (panel a) and (001) (panel b) axes. The scans were collected at $T = 100$ K.}
\end{figure}

Before discussing the implications of the magnetic mode softening, we briefly address the pressure dependence of the magnetic structure of Sr$_3$Ir$_2$O$_7$. Magnetic diffraction peaks are shown in Fig.~\ref{fig:HKL_scan} up to 12 GPa. Spectra are collected by monitoring the elastic signal while scanning along high-symmetry lines of the reciprocal lattice parallel to the (100) and (001) directions (Fig.~\ref{fig:HKL_scan}(a) and \ref{fig:HKL_scan}(b), respectively). As can be seen, the magnetic diffraction peaks are broadened and damped with pressure, but overall the long-range AFM order survives at least until 12 GPa. Above this pressure value, we noticed a degradation of the sample quality, as testified by a drastic broadening and suppression of the charge diffraction peaks. Therefore, results can be safely discussed only up to 12 GPa.

\begin{figure}
	\centering
	\includegraphics[width=\columnwidth]{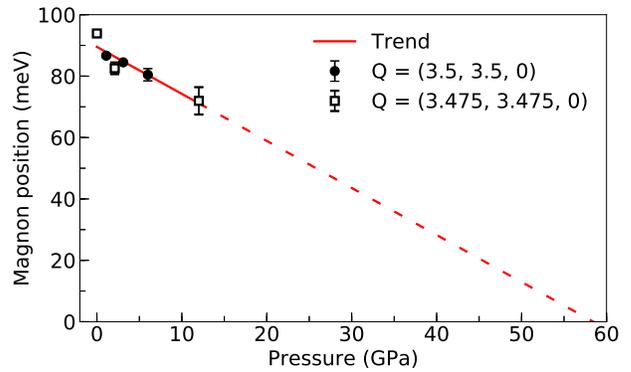}
	\caption{\label{fig:pressure_trend}Energy position of the magnon peak (black circles and squares) as results from the fit of the data and linear extrapolation of the pressure evolution of the magnon softening (red line) until the collapse of the magnetic gap, which is estimated to occur at a pressure of $\sim 58$ GPa.}
\end{figure}

Having established that the underlying magnetic structure is preserved, we now discuss the pressure evolution of the magnetic mode. This is summarized in Fig.~\ref{fig:pressure_trend}, which displays the pressure dependence of the magnon peak position (black circles and squares) together with a linear fit (red solid line) to the experimental data points. The softening rate is $\approx 1.5$ meV/GPa. A linear extrapolation of data points to higher pressures (dashed line) hints towards a collapse of the magnetic mode around $\sim 55$-60 GPa. We note that a structural phase transition has been reported in Sr$_3$Ir$_2$O$_7$ at 54 GPa \cite{Donnerer2016}, therefore suggesting a strong link between magnetic and lattice degrees of freedom, possibly resulting from the strong spin-orbit coupling. 

In relation to the debate about the most suitable theoretical model to describe the magnetic dynamics of Sr$_3$Ir$_2$O$_7$, we recall that high-pressure diffraction measurements revealed a higher compressibility of the $a$ and $b$ axes with respect to the $c$ axis \cite{Donnerer2016}. Indeed, applied pressure is mostly accommodated by rotating the IrO$_6$ octahedra around the $c$ axis, therefore further departing the in-plane Ir-O-Ir bond angle from the straight (\SI{180}{\degree}) geometry. Along the $c$ axis, instead, the Ir-O-Ir bond angle is unchanged, while the Ir-O-Ir bond length is shortened \cite{Donnerer2016,Ding2016}. Since electronic interactions are strongly affected by the bond geometry \cite{Jackeli2009}, the pressure dependence of the dominant magnetic couplings can be guessed. In particular, the intralayer exchange coupling is expected to decrease with pressure, while the interlayer exchange coupling should be less affected by pressure and, if any, it is expected to increase. In light of these expectations, we speculate that it is easier to reconcile our experimental results with the linear spin-wave theory approach \cite{Kim2012a}, because the magnetic gap in the quantum dimer model should scale with the interlayer exchange coupling to first order \cite{MorettiSala2015}. Our qualitative discussion is consistent also with a recent refinement of the crystal structure of Sr$_3$Ir$_2$O$_7$, since the departure from the pure tetragonal lattice is found to be very small \cite{Hogan2016b}. A comprehensive interpretation of the experimental results goes well beyond the scope of the present paper, which is more to establish the general feasibility of performing high-pressure RIXS experiments of magnetic excitations and provide the scientific community with an additional tool to study electronic and magnetic dynamics in matter under extreme conditions.

\section{Conclusions and outlook}

To summarize, we described recent instrumental upgrades carried out at beamline ID20 of the ESRF that allowed us to perform RIXS measurements of magnetic excitations at high pressure. In particular, we measured the magnetic excitations of the bilayer perovskite Sr$_3$Ir$_2$O$_7$ up to 12 GPa and observed a softening of the magnetic mode with pressure at a rate of $\approx 1.5$ meV/GPa. Extrapolation of this behavior at higher pressures shows that the collapse of the magnetic gap might be concomitant with the structural phase transition reported at 54 GPa \cite{Donnerer2016}, suggesting a strong link between magnetic and lattice degrees of freedom. 
Most importantly, we demonstrated the feasibility of RIXS experiments of low-energy excitations at extreme conditions of temperature and pressure.

\begin{acknowledgments} 
We acknowledge the ESRF for providing beamtime and technical support. The authors would like to thank V.~Cerantola, S.~Petitgirard, A.~D.~Rosa and C.~Sahle for fruitful discussions.
\end{acknowledgments}

\bibliography{biblio}

\end{document}